\documentclass[iop]{emulateapj}
\newcommand{\Msun}{\,M_{\odot}}
\newcommand{\Lsun}{\,L_{\odot,r}}
\newcommand\ionn[2]{#1$\;${\scshape{#2}}}
\def\kms{\ {\rm km\, s}^{-1}}
\def\teff{T_{\rm eff}}
\def\logg{{\rm log}\,g}
\def\Re{R_{\rm e}}

\shortauthors{CONROY, VAN DOKKUM, VILLAUME}

\shorttitle{A Super-Salpeter IMF in the center of NGC 1407}

\begin{document}

\title{The Stellar Initial Mass Function in Early-Type Galaxies From
  Absorption Line Spectroscopy. IV.  A Super-Salpeter IMF in the
  center of NGC 1407 from Non-Parametric Models}

\author{Charlie Conroy\altaffilmark{1},
  Pieter G. van Dokkum\altaffilmark{2},
Alexa Villaume\altaffilmark{3}}

\altaffiltext{1}{Department of Astronomy, Harvard University,
  Cambridge, MA, 02138, USA}
\altaffiltext{2}{Department of Astronomy, Yale University, New Haven,
  CT, 06511, USA}
\altaffiltext{3}{Department of Astronomy and Astrophysics, University
  of California, Santa Cruz, CA 95064, USA}

\slugcomment{Submitted to ApJ}

\begin{abstract}

  It is now well-established that the stellar initial mass function
  (IMF) can be determined from the absorption line spectra of old
  stellar systems, and this has been used to measure the IMF and its
  variation across the early-type galaxy population.  Previous work
  focused on measuring the slope of the IMF over one or more stellar
  mass intervals, implicitly assuming that this is a good description
  of the IMF and that the IMF has a universal low-mass cutoff.  In
  this work we consider more flexible IMFs, including two-component
  power-laws with a variable low-mass cutoff and a general
  non-parametric model.  We demonstrate with mock spectra that the
  detailed shape of the IMF can be accurately recovered as long as the
  data quality are high (S/N$\gtrsim300$ \AA$^{-1}$) and cover a wide
  wavelength range ($0.4\mu m -1.0\mu m$).  We apply these flexible
  IMF models to a high S/N spectrum of the center of the massive
  elliptical galaxy NGC 1407. Fitting the spectrum with non-parametric
  IMFs, we find that the IMF in the center shows a continuous rise
  extending toward the hydrogen-burning limit, with a behavior that is
  well-approximated by a power-law with an index of $-2.7$. These
  results provide strong evidence for the existence of extreme
  (super-Salpeter) IMFs in the cores of massive galaxies.

\end{abstract}



\section{Introduction}
\label{s:intro}

The IMF connects a variety of astrophysical phenomena including star
formation, stellar feedback, the relative number of stellar remnants,
and heavy element production, and is a critical ingredient in modeling
the integrated light from distant galaxies.  Precision knowledge of
the IMF is essential for understanding the dark matter content within
galaxies \citep[e.g.,][]{Cappellari06} and measuring supermassive
black hole masses \citep[e.g.,][]{Gebhardt09, McConnell12}.
Unfortunately, the IMF is very difficult to measure when individual
stars cannot be resolved, owing to the intrinsic faintness of lower
main sequence stars relative to the evolved giants
\citep[e.g.,][]{Conroy12a}.

Nonetheless, in recent years there have been many reported
measurements of the IMF in distant galaxies by a variety of
techniques, including kinematic analysis, strong lensing, stellar
population synthesis, and microlensing \citep[e.g.,][]{Cenarro03,
  vanDokkum10, Treu10, Conroy12b, Spiniello12, Cappellari12,
  LaBarbera13, Ferreras13, Conroy13c, McDermid14, Schechter14,
  Posacki15, Lyubenova16}.  The emerging consensus is that the IMF
appears to vary systematically from a Milky Way-like IMF
\citep{Kroupa01, Chabrier03} at lower galaxy masses to a
\citet{Salpeter55} IMF, or even more bottom-heavy IMF, for the most
massive galaxies.  This consensus is not without challenges
\citep[e.g.,][]{Smith14, Smith15, Newman16}.

At the same time, evidence has emerged that the IMF may vary strongly
within galaxies, at least at high masses \citep{Martin-Navarro15,
  LaBarbera16a, vanDokkum16, Davis17}.  Not all authors find evidence
for radial variation in the IMF \citep{McConnell15, Zieleniewski16},
though it is not clear if the different conclusions are due to small
sample sizes or the analysis techniques.  The existence of strong IMF
gradients implies that direct comparison between different techniques
requires careful consideration of aperture effects.

When inferring the IMF from the depth of stellar absorption lines, all
analyses to date have focused on parameterized IMFs, e.g., a single
power-law or a broken power-law, with a fixed low-mass cutoff.
Recently, \citet{Spiniello15} and \citet{Lyubenova16} have combined
stellar population and dynamical constraints for individual galaxies
in order to place stronger constraints on the shape of the IMF.

Here we go one step further and attempt to constrain the general form
of the IMF directly from the absorption line spectra of old stellar
systems.  We consider several models, including a two-part power-law
with a variable low-mass cutoff and a general non-parametric IMF
\cite[see also][who explored the ability to reconstruct the IMF from
mock data]{Dries16}.  Our goals are 1) to assess what is measurable
from absorption line spectra in the idealized limit of perfect models,
and 2) to reconstruct from real data the IMF under different
assumptions and assess how the derived mass-to-light ratio depends on
these assumptions.

\vspace{2cm}


\section{Models \& Spectral Fitting}
\label{s:models}

\subsection{Overview}

Our approach to constructing stellar population models follows
\citet[][CvD12a]{Conroy12a}, with several important updates.  First
and foremost, the models now span a wider range of ages and
metallicities, with ages extending from $1-13.5$ Gyr and metallicities
from [Z/H]=-1.5 to [Z/H]=+0.25.  Second, we now adopt the MIST stellar
isochrones \citep{Choi16} which obviate the need to stitch together
various isochrone databases as done in CvD12a.  Third, we have
obtained IRTF NIR spectra for 283 stars from the MILES spectral
library \citep{Sanchez-Blazquez06}, providing continuous spectral
coverage from $0.35-2.5\mu m$ for stars with well-known stellar
parameters over a wide range in metallicity.  The details of the new
library are presented in \citet{Villaume16}.  In addition to the new
library, we have created a polynomial spectral interpolator that
provides stellar spectra for arbitrary $\teff$, $\logg$, and [Z/H],
the details of which are presented in \citet{Villaume16}.  We include
the library of M dwarfs from \citet{Mann15} to more fully population
the cool dwarf regime for the purposes of constraining the
interpolator. Fourth, the theoretical response functions, which are
computed from the ATLAS and SYNTHE model atmosphere and spectrum
synthesis package \citep{Kurucz70, Kurucz93}, have been recomputed
with updated atomic and molecular data and for stellar parameters
covering the full age and metallicity range of the new models.  The
element response functions are then tabulated as a function of age and
metallicity assuming a Kroupa IMF.  We have checked that the derived
mass-to-light ratios do not change substantially if the response
functions are instead computed with a Salpeter IMF \citep[but see][for
an example where this choice may affect the NIR \ionn{Na}{i} lines at
$>1\mu m$]{LaBarbera16b}.

The methodology for fitting models to data is similar to the approach
described in \citet{Conroy12b} and \citet{Conroy14a}.  Namely, we
employ a Markov chain Monte Carlo algorithm \citep{Foreman-Mackey13}
in order to sample the posteriors of the following parameters:
redshift and velocity dispersion, a two component star formation
history (two bursts with free ages and relative mass contribution),
the overall metallicity, [Z/H], 18 individual elements (though many
are irrelevant from the standpoint of the IMF, e.g., Sr and Ba), the
strengths of five emission line groups (including the Balmer lines
which are assumed to have relative line strengths determined by Case B
recombination), a hot star component with $\teff$ that can vary from
8,000\,K to 30,000\,K, and IMF parameters described below.  In
addition, we fit for several data-related systematics, including a a
multiplicative factor applied to the observed errors, a sky emission
spectrum that is used to increase the errors specifically around sky
lines, and a telluric absorption spectrum, whose overall strength is a
free parameter.  The parameters affecting the observed errors are
given a penalty in the likelihood calculation in order to discourage
the fitter from inflating the errors to be arbitrarily high.  Mismatch
between the observed and model continuum shape is treated by fitting a
polynomial to the ratio of the model and data.  This polynomial is
recomputed for each likelihood evaluation.

\subsection{IMF Models}

In all models in this work the IMF slope above $1.0\Msun$ is assumed
to have the \citet{Salpeter55} value of 2.35, the upper mass limit is
$100\Msun$ and stellar remnants are included in the final mass
calculation following \citet{Conroy09a}.  We explore three options for
the parameterization of the IMF.  The first option has two free
parameters, $\alpha_1$ and $\alpha_2$, corresponding to the
logarithmic slope of the IMF ($dn/dm\propto m^{-\alpha}$) over the
mass intervals $0.08<M/\Msun<0.5$ and $0.5<M/\Msun<1.0$. In this first
option the low-mass cutoff is fixed to $m_c=0.08\Msun$.  Grids of
models were pre-tabulated with $\alpha_i$ ranging from $0.5-3.5$.
Option one is the model IMF adopted in all of our previous work
\citep[e.g.,][]{Conroy12b, Conroy13c, vanDokkum16}.  Option two is the
same as option one except that the low-mass cutoff, $m_c$ is a free
parameter.  Pre-computed grids were made for $0.08<m_c/\Msun<0.40$.

Option three is a non-parametric\footnote{The term ``non-parametric''
  is a something of a misnomer, as there are many parameters in the
  model (one for each bin), in addition to the choice of intra-bin
  behavior.  In the present case, the term is meant to indicate that
  the IMF is not required to follow an analytic (parametric)
  equation.}  IMF.  We have pre-computed what we call ``partial SSPs''
(simple stellar populations), which are SSPs computed within a bin in
stellar mass of width $0.1\Msun$ from $0.2-1.0\Msun$, and
$0.08-0.2\Msun$ for the lowest mass bin.  One must still choose the
IMF weighting within each bin; for this work we adopt a Salpeter IMF
intra-bin weighting because direct measurements of the IMF are
consistent with Salpeter-like slopes, at least for $>0.5\Msun$
\citep{Bastian10}.  We will also consider Kroupa-like intra-bin
weighting to explore the impact of this assumption.  In a more general
treatment the intra-bin weighting would be an additional free
parameter.  In order to keep the number of free parameters manageable,
we only explicitly include five parameters in the model for the bins
starting at 0.08, 0.3, 0.5, 0.7 and 0.9$\Msun$.  The weights for the
other bins are computed via linear interpolation between the fitted
bins.  We emphasize that all bins contribute to the final model and
hence to the $\chi^2$ minimization.  The overall normalization has no
effect on the spectrum nor the mass-to-light ratio, so to further
decrease the number of free parameters we have fixed the weight for
the highest-mass bin to 1.0.  This leaves four free parameters for the
non-parametric IMF option.  Finally, we adopt a regularization
procedure in order to ensure that the final IMF is smooth.  The prior
requires that the logarithmic slope of the IMF between two mass bins
cannot change from negative to positive when moving from lower to
higher masses (e.g., if the logarithmic slope in the $0.2-0.3\Msun$
bin is negative while the slope in the $0.3-0.4\Msun$ bin is positive,
then the prior on that point in parameter space is set to 0.0).

\begin{figure*}[!t]
\center
\includegraphics[width=0.9\textwidth]{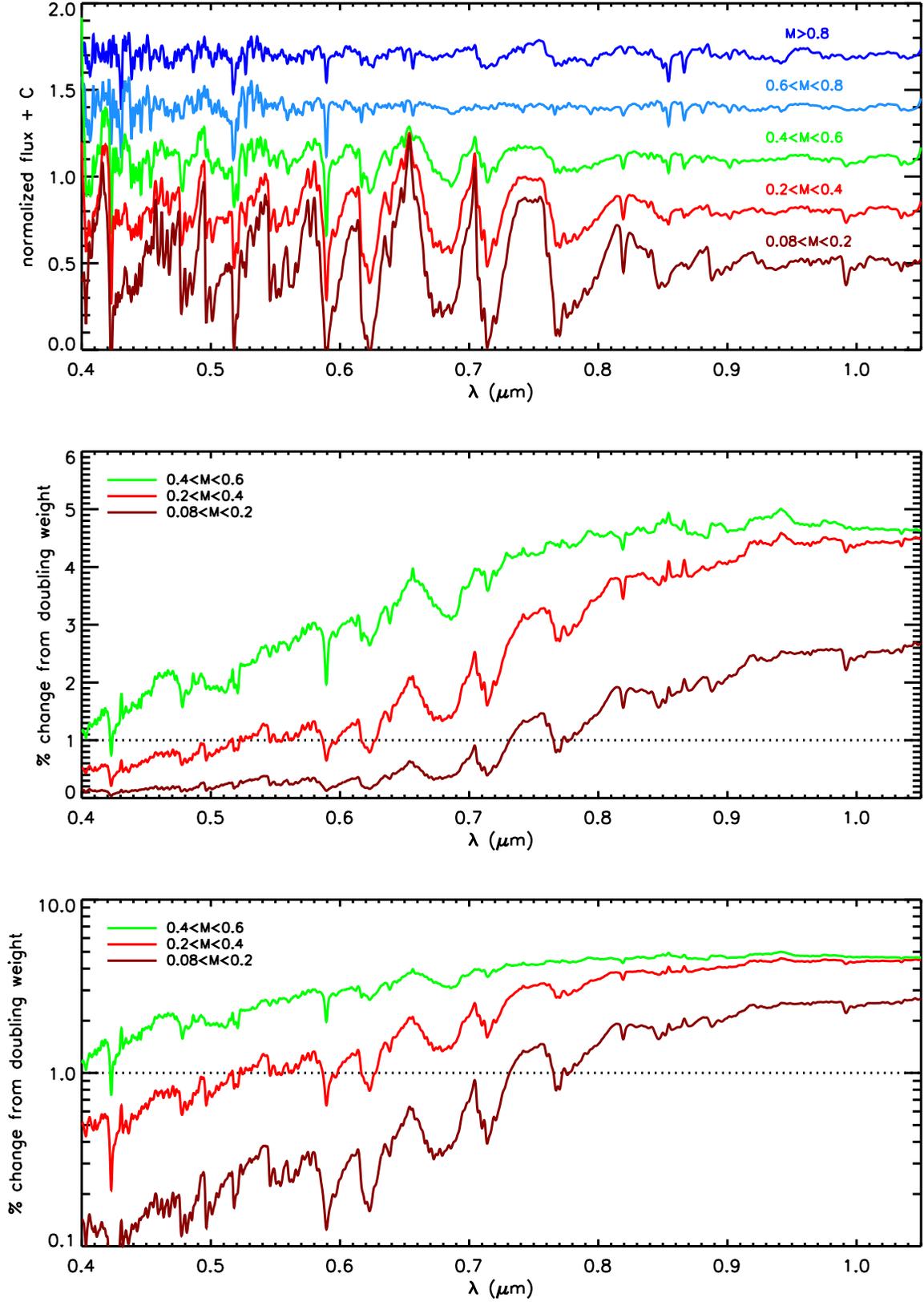}
\vspace{0.1cm}
\caption{Behavior of partial SSPs within five mass bins for
  [Z/H]$\,=0.0$ and an age of 13.5 Gyr.  Models have been smoothed to
  $\sigma=300\kms$.  Top panel: Partial SSPs continuum-normalized by a
  polynomial and vertically offset in increments of 0.3.  Notice that
  the spectrum for each stellar mass bin contains unique spectral
  signatures.  Middle panel: Effect on the combined spectrum from
  doubling the weight assigned to each mass bin below $0.6\Msun$.  The
  reference model assumes a Salpeter IMF.  Bottom panel: Same as the
  middle panel, with a logarithmic y-axis to highlight the behavior in
  the blue.  A dotted line is shown at 1\% the lower panels to guide
  the eye.}
\label{fig:partial}
\end{figure*}

\begin{figure*}[!t]
\center
\includegraphics[width=0.95\textwidth]{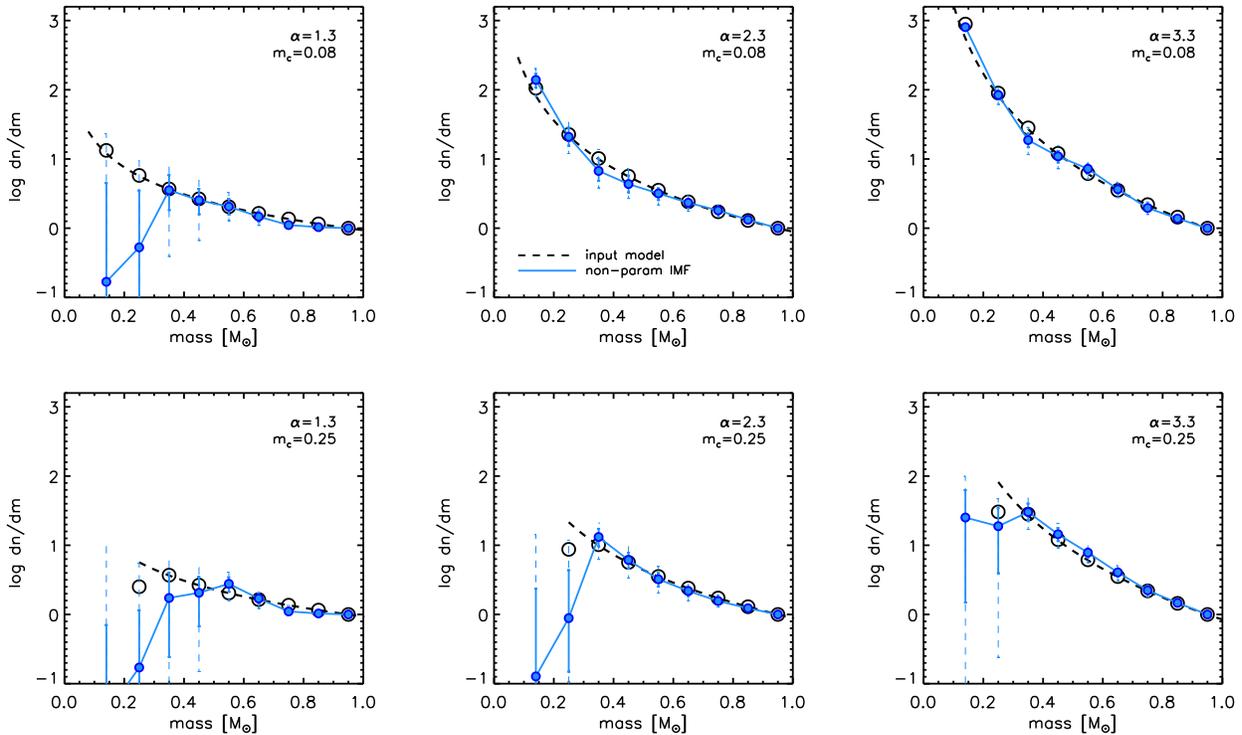}
\vspace{0.1cm}
\caption{Tests of IMF recovery with mock spectra.  Each panel shows
  results for a different input IMF with a single power-law slope,
  $\alpha$, and low-mass cutoff, $m_c$.  The input model is shown
  (dashed lines), as is the input model binned to the mass resolution
  of the non-parametric model (open symbols).  The best-fit
  non-parametric IMFs are shown (solid symbols) along with $16\%-84$\%
  confidence limits (solid lines) and $2.5\%-97.5$\% confidence limits
  (dashed lines).  Note that only five of the nine mass bins are
  independent parameters; the 2nd, 4th, 6th, and 8th bins are
  determined via linear interpolation of the other bins.  The mock
  spectra were generated with S/N$=300$ \AA$^{-1}$.}
\label{fig:mock1}
\end{figure*}

\begin{figure}[!t]
\center
\includegraphics[width=0.48\textwidth]{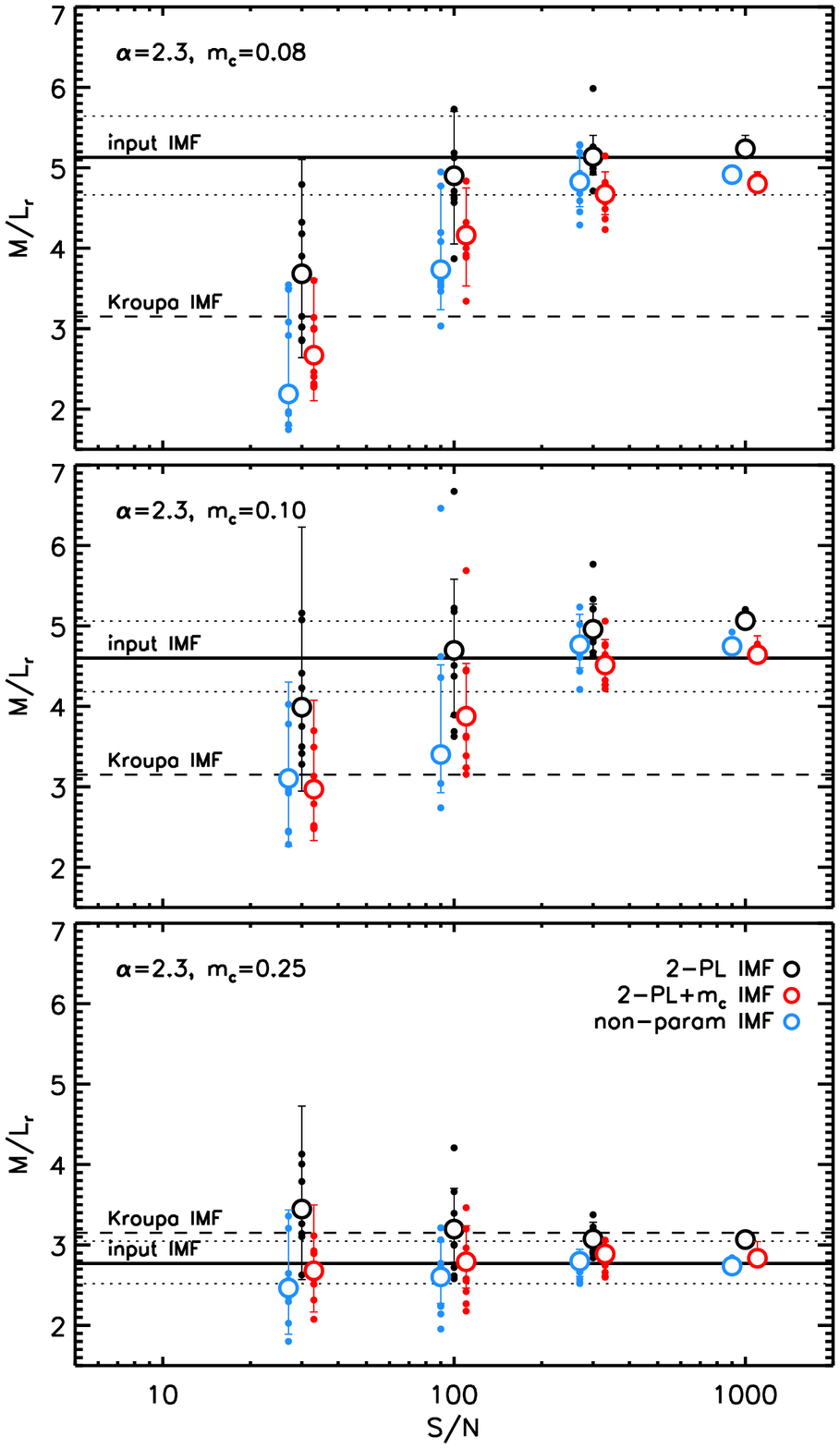}
\vspace{0.1cm}
\caption{Test of mass-to-light ratio recovery with mock spectra.  The
  best-fit $M/L_r$ are shown as a function of the input S/N, which is
  assumed to be constant with wavelength.  Ten independent
  realizations were fit for each S/N value.  In each case the mock
  data were fit with three models: a two-part power-law IMF with fixed
  low mass cutoff (black points), a two-part power-law IMF with
  variable low mass cutoff (red points), and a non-parametric IMF
  (blue points).  Fits are offset slightly along the x-axis for
  clarity.  The large symbols denote the median $M/L_r$ across the 10
  realizations and the error bars represent the median 16\% and 84\%
  confidence limits.  The true $M/L_r$ is shown as a solid line
  ($\pm10$\% is marked by dotted lines), and is compared to an $M/L_r$
  computed for a Kroupa IMF (dashed line).  The panels show results
  for an input spectrum with $\alpha=2.3$ and $m_c=0.08\Msun$ (top),
  $m_c=0.10\Msun$ (middle), and $m_c=0.25\Msun$ (bottom).}
\label{fig:mlr}
\end{figure}

The choice of model priors can be an important, often hidden
assumption when fitting models to data, especially when parameters are
poorly constrained (e.g., when the S/N of the data is low).  In our
case we have adopted uniform, wide priors in all parameters, where the
range is in most cases only limited by the range of the pre-computed
grids.  For the high S/N regime explored in this paper, the priors
play a minimal role.  The variables governing the parameterized IMF
models (options one and two) are allowed to vary over the full range
offered in the pre-computed grids.  For the case of the non-parametric
IMF, the models are parameterized with weights, $w_i$ such that the
IMF within a bin is $w_i\,\int m^{-2.35}dm$.  In these units a
Salpeter IMF would have $w_i=1.0$.  The priors on these weights are
flat in log space from -5.0 to +3.0 (subject to the regularization
condition mentioned above).

Figure \ref{fig:partial} presents an overview of the partial SSPs that
are the building blocks for the non-parametric IMF models.  We have
grouped together the partial SSPs in mass bins $0.2\Msun$ wide, except
for the lowest mass bin, for display purposes.  The models shown in
this figure were computed assuming [Z/H]=0.0 and an age of 13.5 Gyr,
though the qualitative trends are generic.  The top panel shows the
models normalized by a polynomial in order to zoom in on the trends in
the absorption lines with mass.  The sensitivity of various features
to surface gravity and temperature is strong throughout the entire
wavelength range (see also CvD12a).  The middle and bottom panels show
the percentage change to the full SSP model when doubling the weight
assigned to each mass bin, assuming a fiducial Salpeter IMF.
Unsurprisingly, the lowest mass bins contribute very little flux in
the blue, which is why the NIR spectral range is critical for robust
constraints on the low mass IMF.

\vspace{2cm}


\section{Mock Tests}
\label{s:mocks}

In this section we explore the extent to which the IMF and $r-$band
mass-to-light ratio, $M/L_r$, can be recovered from mock data (in this
paper mass-to-light ratios are quoted in units of $\Msun/\Lsun$).  We
create mock spectra with a fixed S/N per \AA\, over the wavelength
range $0.4\mu m-1.015\mu m$ and a velocity dispersion of $300\kms$
(these characteristics were chosen to be similar to the observations
discussed in the following section).  In all cases the age is fixed to
10 Gyr, the metallicity is [Z/H]=0.0, with solar-scaled abundance
patterns, and the various nuisance parameters are set to zero.  The
only component that varies is the functional form of the IMF.  The
input mock spectra are constructed with a parametric IMF with a single
logarithmic slope $\alpha$ over the mass range $m_c-1.0\Msun$ where
$m_c$ is the lower-mass cutoff.  Note that this IMF parameterization
is not, in general, contained within the model space for the
non-parametric models.  This only occurs in the case where
$\alpha=2.3$ and $m_c$ equals one of the mass boundaries for the
non-parametric IMF; in this case the parametric and non-parametric
models exactly agree because the intra-bin weighting for the latter
models assumes $\alpha=2.3$.  These mock spectra therefore offer a
strong test of the ability of the non-parametric models to recover
general IMF behavior.

\begin{figure}[!t]
\center
\includegraphics[width=0.45\textwidth]{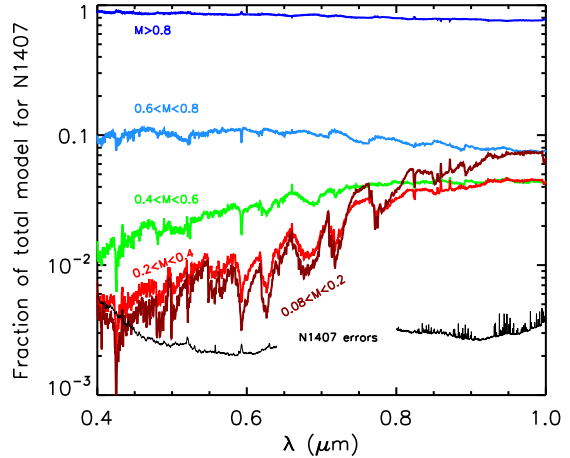}
\vspace{0.1cm}
\caption{Fraction of the total best-fit model spectrum for NGC 1407
  contributed by stars in five stellar mass intervals.  Also shown is
  the error level for the observed central spectrum of NGC 1407.}
\label{fig:n1407_flux}
\end{figure}

\begin{figure}[!t]
\center
\includegraphics[width=0.45\textwidth]{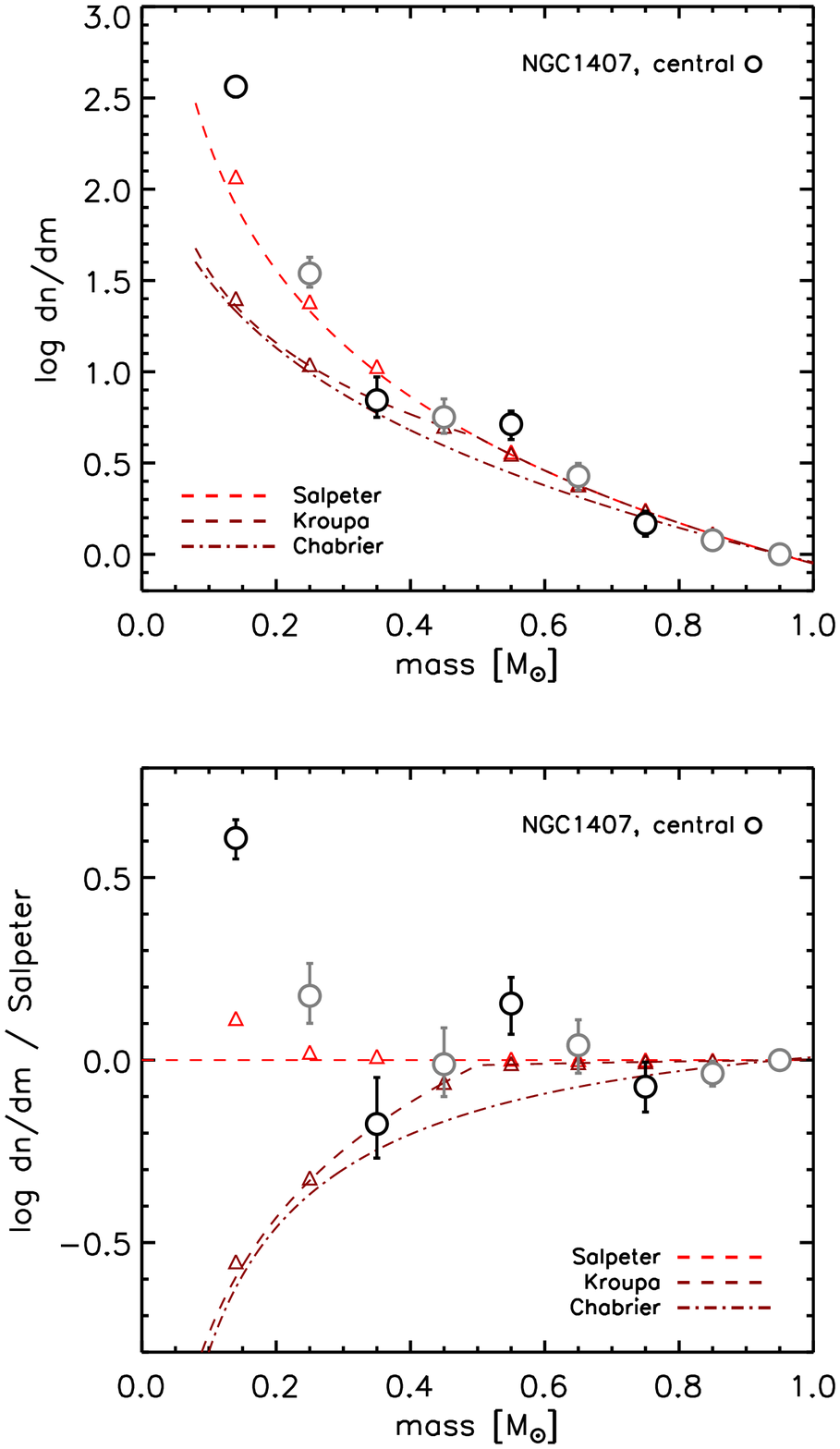}
\vspace{0.1cm}
\caption{Best-fit IMF for the central spectrum of NGC 1407 assuming a
  non-parametric IMF.  Error bars represent 16\%-84\% confidence
  limits.  While all mass bins contribute to the $\chi^2$
  minimization, there are only four independent parameters, indicated
  by black symbols.  Results are compared to Salpeter, Kroupa, and
  Chabrier IMFs (lines) and Salpeter and Kroupa IMFs averaged over the
  stellar mass bins used in the non-parametric analysis (diamonds).
  All IMFs are normalized to unity at $0.95\Msun$.  Top panel shows
  the IMFs while the bottom panel shows the IMFs divided by a Salpeter
  IMF in order to highlight the behavior at low masses. }
\label{fig:n1407inout}
\end{figure}

Figure \ref{fig:mock1} shows the inferred non-parametric IMFs for mock
spectra with six input IMFs.  All of the mock spectra were computed
with S/N$=300$ \AA$^{-1}$. The best-fits (solid symbols) are compared
to the input IMFs (dashed lines) and the input IMF averaged over the
same mass bins as the non-parametric IMF models (open symbols).
Overall, the recovery of the input IMF with the non-parametric models
is remarkable.  In the case of the shallow input IMF ($\alpha=1.3$)
the contribution to the total flux from the lowest mass bins is very
small, and so the non-parametric IMF recovery is biased low, a
consequence of the very wide priors on the individual IMF weights.
However, the very fact that these low-mass bins contribute very little
to the light and so are difficult to constrain also implies that they
contribute little to the mass.  As an example, the bias shown in the
upper left panel results in only a 16\% bias in the recovered $M/L_r$

We explore the $M/L_r$ ratios explicitly in Figure \ref{fig:mlr}.
This figure shows the best-fit $M/L_r$ as a function of S/N for mock
spectra constructed with an IMF with a Salpeter slope and a low-mass
cutoff of $0.08\Msun$ (top panel), $0.10\Msun$ (middle panel) and
$0.25\Msun$ (bottom panel).  Each mock spectrum was fit three times
for the three IMF options described in Section \ref{s:models}.  For
each S/N ten mock spectra were created with identical parameters and
independent realizations of the noise.  Results from fitting each
realization are shown as small symbols and the median is shown as
large open symbols.  Error bars denote the median $16\%-84\%$
confidence limits and should be interpreted as the ``typical'' error
on a single measurement.  The $M/L_r$ for a \citet{Kroupa01} IMF is
indicated by the dashed lines.

Although the overall agreement is good, it is clear from this figure
that there can be important biases in the recovered $M/L_r$ depending
on the S/N, the adopted functional form of the fitted IMF, and the
input IMF.  Results at S/N$<100$ tend to be biased low when the
low-mass cutoff is $<0.25\Msun$.  This occurs because the data are not
of sufficient quality to constrain the low-mass behavior, and so the
posteriors tend toward the priors, which in our setup favors low
mass-to-light ratios.  For S/N$\ge300$ \AA$^{-1}$ the results are
never biased by more than 10\% for the input models explored here.

Another important conclusion to draw from Figure \ref{fig:mlr} is that
the IMF `mis-match' parameter, $\alpha_{\rm IMF}\equiv (M/L) /
(M/L)_{\rm MW}$, is a good summary statistic when S/N$\ge300$
\AA$^{-1}$, irrespective of the detailed IMF shape (although we have
not tested single power-law models in this paper).  This is important
and suggests that the $\alpha_{\rm IMF}$ parameter is a reliable,
compact metric for IMF measurements.

The conclusion from this section is that the full, non-parametric IMF
can be robustly inferred from the spectrum, given sufficiently high S/N
data.  For the specific case of a 10 Gyr solar metallicity population
with a dispersion of $\sigma=300\kms$ the requirement is
S/N$\gtrsim300$ \AA$^{-1}$ if the IMF is Salpeter-like and extends to
$0.08\Msun$; a shallower IMF and/or one with a higher-mass cutoff will
have less severe S/N limitations.  This requirement will depend, at a
minimum, on the velocity dispersion, age, and metallicity of the
population, and the wavelength range of the data.  Attempts to measure
the IMF from fitting absorption line spectra should therefore include
tests with mock observations to assess potential biases in the derived
parameters.

\begin{figure}[!t]
\center
\includegraphics[width=0.45\textwidth]{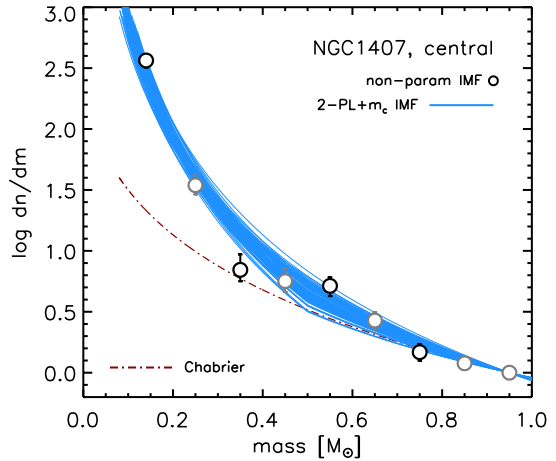}
\vspace{0.1cm}
\caption{Best-fit IMF for NGC 1407 comparing two choices for the
  underlying IMF model.  Open symbols represent the non-parametric
  model, and the blue lines represent 100 random draws from the
  posteriors of a three parameter model that includes a two-part
  power-law below $1\Msun$ and a free low-mass cutoff.  A
  \citet{Chabrier03} IMF is shown for comparison.}
\label{fig:n1407compare}
\end{figure}


\section{Results}
\label{s:res}

We now turn to applying these models to observations.  Here we focus
on the massive elliptical galaxy NGC 1407 ($D=23$ Mpc, $\Re=76\arcsec
= 8.5$ kpc).  We recently observed this galaxy with the LRIS
instrument on the Keck I telescope, obtaining high S/N spectra as a
function of radius.  The data extend beyond $1\,\Re$ although the S/N
in the red drops below 200 at $\approx0.4\,\Re$.  Details of the
observing strategy and data reduction can be found in
\citet{vanDokkum16}.  Here we fit the data over the wavelength
intervals $0.4\mu m- 0.64\mu m$, $0.80\mu m- 0.892\mu m$, and
$0.963\mu m-1.015\mu m$.  The gap at $0.64\mu m- 0.80\mu m$ is due to
the dichroic and grating tilt, while the gap at $0.892\mu m-0.963\mu m$
reflects the fact that atmospheric absorption strongly contaminates
that spectral region.  We focus on the central region within
$\pm0.33\arcsec$.  The detailed stellar population gradients,
including $M/L_r$ and IMF gradients (based on simpler IMF models), are
presented in \citet{vanDokkum16}.

We begin with Figure \ref{fig:n1407_flux}, where we show the
best-fitting model for NGC 1407.  The model is displayed as fractional
contributions to the total model spectrum for stars in five mass
intervals.  For comparison we also include the fractional errors for
the observed spectrum of NGC 1407.  Not surprisingly, the most massive
stars dominate the flux, contributing $\gtrsim90$\% of the flux over
the entire wavelength range.  It is noteworthy that each remaining
mass bin interval contributes at least 4\% per pixel over wavelength
intervals of at least 1000\AA.  Compared to the fractional errors on
the data, which are at the level of $\approx 3\times10^{-3}$, it is
clear that each mass interval is well constrained by the data.

Figure \ref{fig:n1407inout} shows the best-fit non-parametric IMF for
the center of NGC 1407.  The top panel shows the IMF while the bottom
panel shows the IMF divided by a Salpeter IMF ($m^{-2.35}$) in order
to highlight the behavior at lower masses.  For comparison we also
show \citet{Salpeter55}, \citet{Kroupa01}, and \citet{Chabrier03} IMFs
(dashed and dot-dashed lines) and those IMFs averaged over the
non-parametric IMF mass bins (diamonds).  Recall that only four bins
are explicitly free parameters (those with bins starting at 0.08, 0.3,
0.5, and $0.7\Msun$); the $0.9-1.0\Msun$ bin is fixed to 1.0 and the
weights for the four remaining bins are determined by linear
interpolation between the other bins.

This figure shows that the central region of NGC 1407 prefers an IMF
considerably heavier at low masses than Salpeter, by a factor of
$\approx2.5$ in the lowest mass bin.  The inferred mass-to-light ratio
is $M/L_r=12.2^{+0.83}_{-0.82}$, or a factor of two (three) times
higher than the mass-to-light ratio for a Salpeter (Kroupa) IMF
(errors on $M/L_r$ reflect the 16\% and 84\% confidence limits).  It
is remarkable that the lowest mass bin is so well-constrained
(although not surprising in light of Figure \ref{fig:partial}).

We have also fit the spectrum of NGC 1407 assuming a three parameter
IMF model (two-part power-law and a variable low-mass cutoff) to
explore the sensitivity of the derived mass-to-light ratio to the
assumed IMF with real data.  The comparison of this model to the
non-parametric IMF is shown in Figure \ref{fig:n1407compare}.  In this
figure we draw 100 random samples from the posteriors of the three
parameter IMF.  The agreement is impressive and lends further weight
to the robustness of the derived IMF.  The best-fit IMF parameters in
this case are $\alpha_1=3.24^{+0.16}_{-0.18}$,
$\alpha_2=2.37^{+0.22}_{-0.22}$, and $m_c=0.10^{+0.01}_{-0.01}$.  The
inferred $M/L_r=11.6^{+1.47}_{-0.94}$ for the three parameter model
agrees very well with the non-parametric IMF-based value.  Moreover,
the low-mass cutoff is constrained to be lower than $0.13\Msun$ at
97.5\% CL, confirming that the IMF for the central region of NGC 1407
remains steep toward the hydrogen-burning limit.  All of the other
derived parameters (age, metallicity, abundance pattern, etc.) agree to
within 0.01 dex or better when comparing results for the two IMF
parameterizations.

We have tested the impact of several model assumptions on these
results.  Recall that the element response functions were tabulated
assuming a Kroupa IMF.  We have re-fit the central spectrum of NGC
1407 with response functions tabulated assuming a Salpeter IMF in
order to assess the impact of this assumption \citep[see
also][]{LaBarbera16b}.  The best-fit $M/L_r$ in this case is higher by
17\%, suggesting that the IMF derived with the fiducial response
functions is slightly lower than the true value.  Moreover, the
$0.82\mu m$ \ionn{Na}{i} feature is slightly better fit when using
response functions constructed with a Salpeter IMF, even though the
best-fit [Na/Fe] abundance is virtually identical in the two cases.
In the future we will explore this issue in greater depth and consider
options for treating the IMF in the response functions in a
self-consistent manner.

We have also explored the sensitivity of these results to the
intra-bin weighting used to compute the partial SSPs.  Our default
models employ Salpeter intra-bin weights.  We have created an
additional set of models with Kroupa-like intra-bin weighting
(power-law indices of $-1.3$ for $M<0.5\Msun$ and $-2.3$ for
$M>0.5\Msun$).  The resulting IMF constraints are nearly identical to
the default model, and the inferred mass-to-light ratio is
$M/L_r=12.1^{+0.86}_{-0.70}$, also nearly identical to the
mass-to-light ratio inferred with the default models.  We conclude
that our results are insensitive to the choice of intra-bin weighting,
at least within the Kroupa-Salpeter range.

\begin{figure}[!t]
\center
\includegraphics[width=0.45\textwidth]{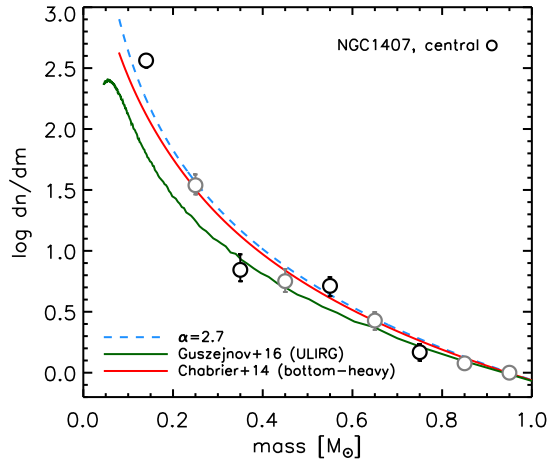}
\vspace{0.1cm}
\caption{Best-fit IMF for NGC 1407 in the central region compared to
  models.  The dashed line is a single power-law IMF of the form
  $m^{-2.7}$, while the solid lines are models from the literature:
  the red line is the bottom-heavy IMF from the formalism of
  \citet[][their Case 3]{Chabrier14} and the green line is the ULIRG
  IMF from \citet{Guszejnov16}.}
\label{fig:n1407in}
\end{figure}


\section{Discussion}

In Paper III of this series \citep{vanDokkum16} we explored stellar
population gradients in six galaxies, including NGC 1407.  Using
simpler IMF models (a two-part power-law with a fixed low-mass cutoff)
we found strong IMF gradients in all six galaxies.  The extreme IMF in
the center of NGC 1407 reported here, derived with more flexible IMF
models, is evidently a phenomenon confined to the inner regions.

Several distinct analytical models have been proposed to explain the
origin of the IMF and its dependence on the properties of the
interstellar medium and/or the galaxy as a whole.  The model proposed
by \citet{Krumholz11} argues that the characteristic fragmentation
mass in a molecular cloud is governed by radiation released by the
collapsing protostar.  This process produces a weak relation between
the characteristic stellar mass and the pressure of the surrounding
gas such that high pressure environments produce lower characteristic
masses.  \citet{Hennebelle08}, \citet{Hopkins12a}, and
\citet{Chabrier14} argue that the IMF emerges from the combined
effects of gravity and lognormal density fluctuations in a turbulent
medium (with the important assumption that the ratio between core and
stellar mass is constant).  Recently, \citet{Guszejnov16} combined
these ideas by considering a model that includes both
gravito-turbulent fragmentation and stellar radiation feedback.

In Figure \ref{fig:n1407in} we compare several of these semi-analytic
models to our recovered IMF for the central region of NGC 1407.  We
include the model for extreme starburst conditions from \citet[][Case
3]{Chabrier14}, and the ULRIG model from \citet{Guszejnov16}.  For
comparison we also include a power-law IMF of the form $m^{-2.7}$, and
note that in the formalism of Chabrier et al. the most extreme IMFs
saturate at an index of $\approx-2.7$.  The agreement between the
observations and the Chabrier et al. model is remarkable given that
there was no fine tuning in the model parameters.

The derived IMF for the central region of NGC 1407 is very steep,
which means that the total stellar mass is sensitive to the adopted
low-mass IMF cutoff.  In our non-parametric analysis, the lowest-mass
bin covers the range $0.08\Msun-0.2\Msun$, so we cannot make strong
statements regarding the cutoff mass, only that it must extend below
$0.2\Msun$.  When computing mass-to-light ratios we have assumed that
the lowest mass bin extends to $0.08\Msun$.  To illustrate the
sensitivity of the total mass to the cutoff, for a single power-law
with $\alpha=2.7$, the mass-to-light ratio is 70\% higher if the
cutoff is $0.05\Msun$ compared to $0.08\Msun$.

It is therefore critical to understand which physical processes set
the low-mass cutoff, and how we might hope to constrain the location
of the cutoff in early-type galaxies.  \citet{Chabrier14} argue, based
on calculations presented in \citet{Masunaga99}, that the minimum mass
for fragmentation may be much higher in early-type galaxies compared
to the Milky Way, owing to the higher temperatures, densities, and
opacities in the former.  For reasonable conditions, the minimum mass
could be comparable to the hydrogen-burning limit, implying a sharp
truncation in the IMF near $0.1\Msun$.

A promising path forward to constrain the low-mass cutoff in
early-type galaxies is the joint modeling of stellar populations,
dynamics, and/or gravitational lensing \citep[e.g.,][]{Barnabe13,
  Spiniello15}.  In this approach, the stellar population analysis
constrains the IMF over the stellar range (as in this work), and the
dynamics and/or lensing probe the integral of the IMF down to the
low-mass cutoff, irrespective of whether that cutoff is in the stellar
or sub-stellar regime.  With this approach one must still make an
assumption regarding the mass in stellar remnants, i.e., the shape of
the IMF above $1\Msun$, if one wants to constrain the low-mass cutoff.
Early work along these lines by \citet{Barnabe13} and
\citet{Spiniello15} appears promising, and suggests that the cutoff in
massive early-type galaxies is indeed near $0.1\Msun$.

In this work we used updated stellar population models to constrain
the detailed shape of the IMF over the range $0.08-1.0\Msun$ in the
center of the massive early-type galaxy NGC 1407.  The IMF in the
galaxy core is very steep, consistent with ${\rm d}n/{\rm d}m\propto
m^{-2.7}$ down to the hydrogen-burning limit.  Such a steep IMF is
also consistent with the IMF predicted for extreme starburst
conditions from \citet{Chabrier14}.  These results demonstrate that it
is possible to directly probe the ``bottom'' of the IMF from
absorption line spectra of old stellar systems.


\acknowledgments 

C.C. thanks Andrew Newman and Paul Schechter for conversations that
lead to the development of these flexible IMF models.
C.C. acknowledges support from NASA grant NNX15AK14G, NSF grant
AST-1313280, and the Packard Foundation.  A.V. acknowledges the
support of an NSF Graduate Research Fellowship.  We thank the referee
for comments that improved the quality of the manuscript.  These
results are based on data obtained with the W. M. Keck Observatory, on
Mauna Kea, Hawaii. The authors wish to recognize and acknowledge the
very significant cultural role and reverence that the summit of Mauna
Kea has always had within the indigenous Hawaiian community. We are
most fortunate to have the opportunity to conduct observations from
this mountain.


\end{document}